\documentclass[12pt,a4paper]{article}
\usepackage[utf8]{inputenc}
\usepackage{amsmath}
\usepackage{amsfonts}
\usepackage{amssymb}
\usepackage{graphicx}
\usepackage{natbib}

\title{\bf Photometric observations of the mutual phenomena of the Galilean satellites at the Pulkovo observatory in 2014-2015}

\author{M.Yu. Khovritchev$^{1}$\footnote{e-mail: deimos@gao.spb.ru}, I.S. Izmailov$^{1}$,\\ E.A.Roshchina$^{1}$, D.L.Gorshanov$^{1}$, A.M. Kulikova$^{1}$\\ 
{\small $^1$Pulkovo Observatory, Russian Academy of Sciences},\\
{\small Pulkovskoe sh. 65, St. Petersburg, 196140 Russia}}

\begin{document}
\maketitle

\begin{abstract}
We present the results of photometric observations of the mutual phenomena in the system of Galilean satellites obtained during 2014-2015. The observations were performed using the 26-inch refractor, Normal Astrograph, ZA-320 telescope of the Pulkovo Observatory (084) and MTM-500 telescope at Pulkovo mountain station at Kislovodsk (C20). We made observations a total of 72 phenomena. We had derived 51 light curves of good and medium quality for 34 events. The RMS of determining the brightness is within a range from 0.02 to 0.19 mag, the average RMS is 0.06 mag. This work was supported by RFBR grant (project 15-02-03025).
\end{abstract}

\section*{Introduction}
Dynamical studies of the Galilean satellites require the series of accurate positional observations. An analysis of the light curves of satellites obtained during the mutual phenomena allows us to get high-precise astrometric data - differences of topocentric (in the case of mutual occultations) or heliocentric (for mutual eclipses) coordinates of two satellites involved in the phenomenon. Mutual phenomena in the system of Galilean satellites of Jupiter are repeated with a period of six years and nine months. The aim of this work is providing observational data for further analysis. Pulkovo Observatory was taking part in the previous international campaigns of mutual phenomena observations in 1995, 1997, 2003 and 2009~(\cite{Emel'yanov_et_al_2011}). Pulkovo Observatory had collect about of ten percents of all world observations during the campaign of 2003~(\cite{Arlot_et_al_2009}). Recent set of phenomena was available for observations at Pulkovo (084) and at Mountain station near Kislovodsk (C20) from October 2014 to August 2015. Jupiter's altitude reached to 45 degrees in the meridian. Ephemeris of mutual phenomena were provided by Celestial Mechanics Department of SAI MSU and presented at the ephemeris server MULTI-SAT~(\cite{Emel'yanov_and_Arlot_2008}).

\begin{table}
\caption{Telescopes, cameras and observers.}
\label{Tab1}
\smallskip

\begin{tabular}{p{0.12\linewidth}|p{0.25\linewidth}|p{0.23\linewidth}|p{0.24\linewidth}|p{0.05\linewidth}}
\hline
Site and code & Telescope & Camera  & Observers & Number of light curves\\
\hline
Pulkovo 084  & 26-inch refractor,\newline D=650~mm,\newline F=10413~mm & FLIProline~9000,\newline FOV $12'\times12'$, \newline scale=$0.24''$/pix  & Izmailov~I.S. & 27\\\hline
Pulkovo 084  & Normal \newline Astrograph,\newline D=330~mm, \newline F=3500~mm & SBIG~STL-11K, \newline FOV $35'\times23'$,\newline scale=0.53$''$/pix  &  Khovrichev~M.Yu., Kulikova~A.M. & 8\\\hline
Pulkovo 084  & ZA-320,\newline D=320~mm,\newline F=3200~mm & SBIG~STL-16803, \newline FOV $39'\times39'$, \newline scale=0.6$''$/pix  & Gorshanov~D.L., Petrova~S.N., Slesarenko~V.Yu., Naumov~K.N., Ivanov~A.V., Sokov~E.N., Kupriyanov~V.V. & 9\\\hline
{\small Mountain Station} {\tiny(Kislovodsk)} C20 & MTM-500M,\newline D=500~mm,\newline F=6520~mm & SBIG~STL-1001E, \newline FOV $21'\times21'$, \newline scale=1.2$''$/pix  & Lyashenko~A.Yu., Rusov~S.A.& 7\\
\hline
\end{tabular}
\end{table}

\begin{figure}
\center{\includegraphics[width=1\linewidth]{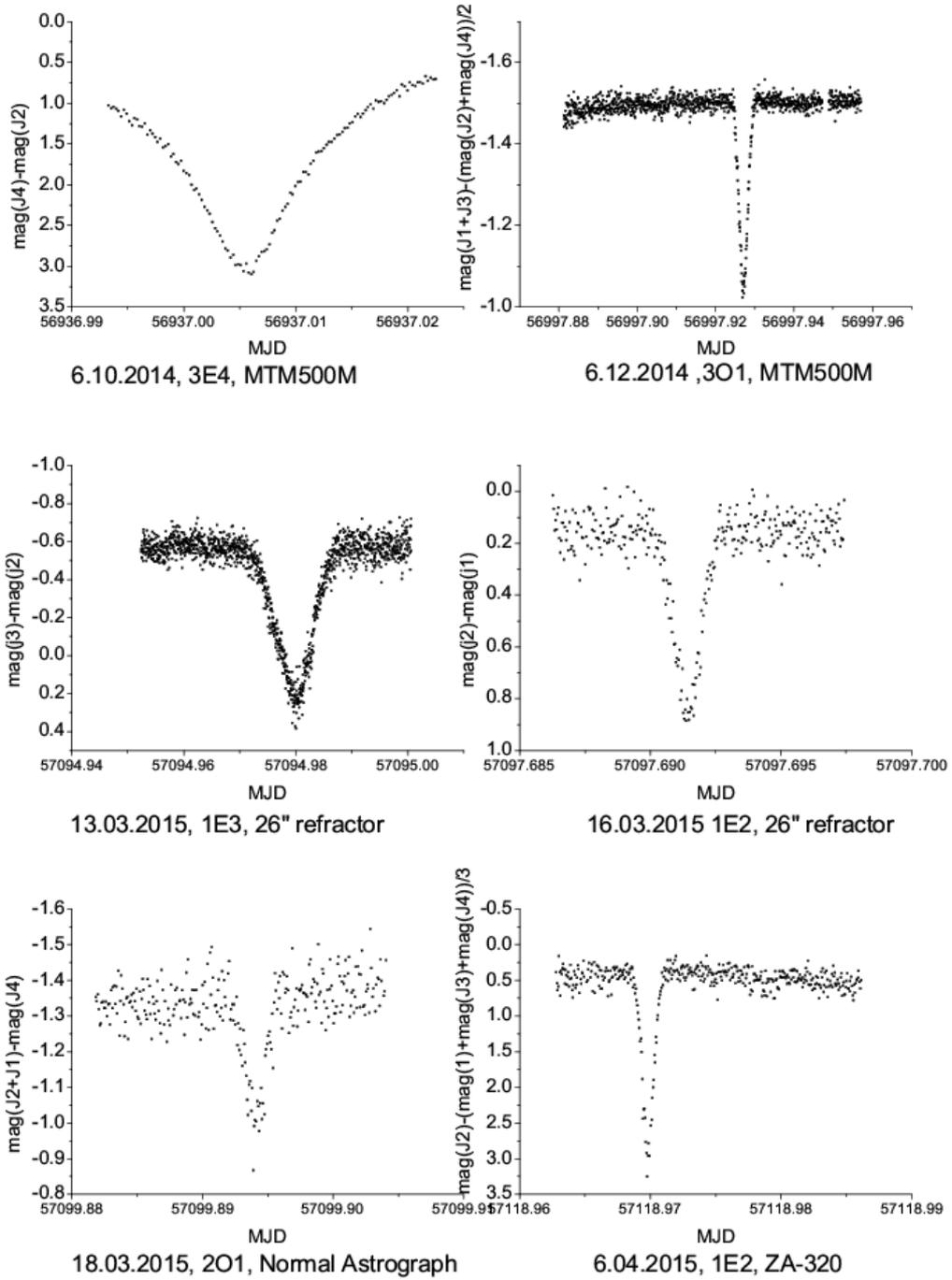}}
\caption{Typical light curves of several phenomena.}
\label{Fig1}
\end{figure}

\section{Observations and data processing}
Observations were carried out with three Pulkovo telescopes (26-inch refractor, Normal Astrograph, astrograph ZA-320) and the MTM-500M telescope of the Pulkovo Mountain Station near Kislovodsk (See Table~\ref{Tab1}). The observations were performed according to the recommendations given in Emel'yanov's paper~(\cite{Emel'yanov_2008}). Observations began in advance before the start of phenomenon and continued for up to the duration time after it's end. The exposure time was varied from 0.1~s to 1~s, depending on weather conditions and the magnitudes of satellites. No filter was used. These CCD images were analyzed and integrated fluxes were calculated for satellite images at the central moment of each frame. Dark current and flat fields were taken into account by the standard way. As a result, we obtained the ratio of satellite light flux and reference object or the difference between magnitudes of the satellite and the reference object. We chose the satellite, which did not involved in a phenomenon, or a background star as a reference object. To correct the background gradient due to planet's scattered light we calculated the background intensities using a linear
 model according to expression $I(x,y)=Ax+By+C$. The parameters $A$, $B$ and $C$ were determined by the least squares method using intensity value of the pixels ($I(x,y)$) within the ring-shaped area around the image of the satellite. Whenever possible, the Jupiter was taken out of the field of view of the camera to minimize errors in determining the background. The least radius of aperture was selected.

\section{The results and conclusions}
In a total, we performed 73 observations of 37 mutual phenomena. Fifty-series of CCD frames of suitable quality and light curves were obtained. Some examples of light curves are shown in Fig.~\ref{Fig1}. The values of magnitude differences were approximated by a quadratic polynomials to estimate preliminary values of photometric precision. The standard deviations of determining the magnitudes are ranged from 0.02 to 0.19 mag. The obtained light curves are available at the Pulkovo database  http://puldb.ru/photometry/phemu2014-2015/.

\section*{Acknowledgments}
This work was supported by Russian Foundation for Basic Research (project no. 15-02-03025).

\end{document}